\documentstyle[epsf]{elsart}
\newcommand{\beq}{\begin{equation}}
\newcommand{\eeq}{\end{equation}}
\newcommand{\beqar}{\begin{eqnarray}}
\newcommand{\eeqar}{\end{eqnarray}}
\newcommand{\ds}{\displaystyle}
\begin{document}
\begin{frontmatter}

\title  {
Elliptic flow at collider energies and cascade string models: 
The role of hard processes and multi-Pomeron exchanges
         }
\author {
E.E.~Zabrodin$^{a,b}$, C.~Fuchs$^{a}$, L.V.~Bravina$^{a,b}$, 
and Amand~Faessler$^{a}$
  } \\
\address{
$^a$Institute for Theoretical Physics, University of T\"ubingen,
    Auf der Morgenstelle 14, D-72076 T\"ubingen, Germany\\
$^b$Institute for Nuclear Physics, Moscow State University,
    RU-119899 Moscow, Russia
         }

\maketitle

\begin{abstract}
Centrality, rapidity, and transverse momentum dependence of hadron
elliptic flow is studied in Au+Au collisions at BNL RHIC energies
within the microscopic quark-gluon string model. The QGSM predictions 
coincide well with the experimental data at $\sqrt{s}=130${\it A\/} 
GeV. Further investigations reveal that multi-Pomeron exchanges and 
hard gluon-gluon scattering in primary collisions, accompanied by the 
rescattering of hadrons in spatially anisotropic system, are the key 
processes needed for an adequate description of the data. These 
processes become essentially important for heavy-ion collisions at 
full RHIC energy $\sqrt{s}=200${\it A\/} GeV. 

\end{abstract}

\end{frontmatter}

\newpage

One of the main goals of experiments on heavy-ion accelerators,
including the Relativistic Heavy-Ion Collider (RHIC) at BNL,
which is operating since 1999, and the coming Large Hadron Collider
(LHC) at CERN, is to determine the equation of state (EOS) of
nuclear matter under extreme conditions and to probe the
deconfinement phase transition to the new state of matter dubbed
quark-gluon plasma (QGP) (see, e.g., \cite{QM99}).
It has been understood long ago that nuclear matter expanding in 
the direction perpendicular to the beam axis of the colliding nuclei
can carry information about the early stage of the reaction and
subsequent evolution of the system \cite{SMG74}. This phenomenon
known as ``collective flow" has been intensively studied both 
theoretically and experimentally during the last two decades 
(see \cite{StGr86,ReRi97,HWW99} and references therein).
Nowadays the expansion in Fourier series is usually applied to
study the azimuthal distribution of particles \cite{VoZh96,PoVo98}:
\beq
\ds
E \frac{d^3 N}{d^3 p} = \frac{1}{\pi} \frac{d^2 N}{dp_t^2 dy} \left[ 
1 + \sum_{n=1}^{\infty} 2 v_n \cos(n\phi) \right] ,
\label{eq1}
\eeq
where $\phi$ is the azimuthal angle, and $p_t$ and $y$ is the 
transverse momentum and the rapidity of a particle, respectively. 
The first term in square brackets represents the isotropic radial 
flow, while the others are referred to anisotropic flow. The 
first Fourier coefficient $v_1 = \langle \cos{\phi} \rangle \equiv
\langle p_x / p_t \rangle $ is called directed flow, and the 
second one $v_2 = \cos{(2 \phi)} \rangle \equiv \langle (p_x /
p_t)^2 - (p_y / p_t)^2 \rangle $ is dubbed elliptic flow.
Here the beam axis is labelled as $z$-axis. Together with the
impact parameter $x$-axis, it defines the reaction plane
perpendicular to the $y$-axis. To study centrality, (pseudo)rapidity,
and transverse momentum dependence of the anisotropic flow the
following triple differential distributions integrated over any two
of variables in given intervals are useful
\beq \ds
v_n(x_i) = \int_{x_j^{(1)}}^{x_j^{(2)}} \cos(n\phi) 
\frac{d^3 N}{dx_j} dx_{j \neq i} \left/ \int_{x_j^{(1)}}^{x_j^{(2)}} 
\frac{d^3 N}{dx_j} dx_{j \neq i} \right. \ ,
\label{eq2}
\eeq
where the variables $x_j,\ j=1,2,3$ denote the impact parameter $b$, 
transverse momentum $p_t$, and (pseudo)rapidity $(\eta)y$.

The importance of elliptic flow as a probe of the hot and dense 
nuclear phase at high energies has been discussed first in 
Ref.~\cite{Olli92}. Both microscopic and macroscopic calculations seem 
to support the idea that elliptic flow is developed at the very early 
stage of a nuclear collision
\cite{Sorplb97,Olli98,LPX99,ZGK99,KSH99,HeLe99,Brac00}. 
Several features, such as a characteristic ``kinky" structure in the 
excitation function of elliptic flow \cite{Sorprl99}, have been 
proposed to search for the QGP phase. The first data on elliptic flow 
of charged particles in the midrapidity range of Au+Au collisions at 
$\sqrt{s} = 130${\it A} GeV, measured by the STAR Collaboration, 
became available recently \cite{ellST00}. It appears that the 
microscopic transport cascade models, e.g. ultra-relativistic quantum 
molecular dynamics (UrQMD) \cite{urqmd} or RQMD \cite{Sorprl99}, 
predict too weak values of $v_2(b, \eta)$ \cite{BlSt00,SPV99},
while hydrodynamic models overpredict the measured 
elliptic flow by about 20-50 \% \cite{ellST00}.
In hydrodynamics the EOS can be softened by the introduction of a
QGP phase and hadron resonance-rich matter \cite{TLS00,KHHH00}.
In the present paper we demonstrate that the elliptic flow at RHIC
can quantitatively be described within cascade string models by the 
mechanism of string excitations due to colour exchange. This procedure 
is different from the FRITIOF model of longitudinal excitations of 
strings \cite{fritiof} which is the basic part of the string sector in
several ultra-relativistic transport models.
For the further analysis the quark-gluon string model (QGSM)
\cite{qgsm} is employed.

\begin{figure}[htp]
\centerline{\epsfysize=16cm \epsfbox{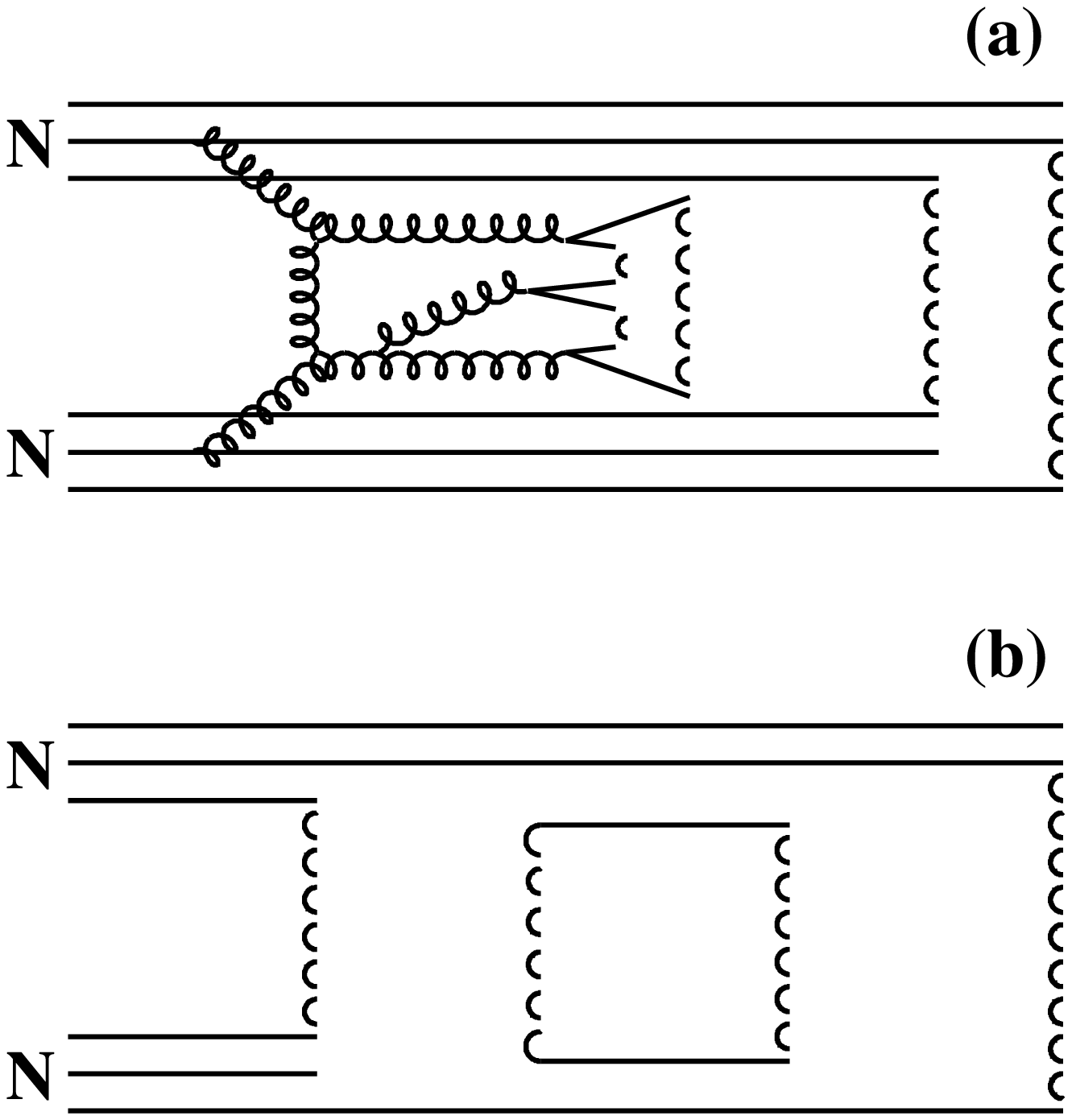}}
\vspace{0.5cm}
\caption{
(a) String formation in hard gluon-gluon scattering and soft
Pomeron exchange in nucleon-nucleon (NN) collision.
(b) Formation of four strings as a result of double Pomeron 
exchange in NN collision.
}
\label{fig1}
\end{figure}

In the hadron-hadron ({\it hh\/}) collision part of the QGSM 
statistical weights, hadron structure functions, and leading quark 
fragmentation functions have been obtained in Ref. \cite{Kaid} within 
the Gribov-Regge theory (GRT) \cite{grt}. This enables one to choose 
subprocesses of string excitations, calculate mass and momentum 
of a string, and simulate the string fragmentation into hadrons
properly. Since a hadron is represented by a vector in the Fock space
of constituents, namely valence and sea quarks, diquarks, gluons,
and their antistates, the strings produced in hadronic collisions
originate from different interactions between the constituents.
Scattering of strings is approximated by scattering of the valence 
quarks and diquarks at the end of the strings, and of hadrons that
are produced in string decays. Similar to the RQMD model
the valence quarks and diquarks are allowed to interact promptly,
while the newly produced hadrons can scatter only after a formation
time related to the time needed to break a string. Together with
subprocesses with quark annihilation and exchange associated with 
the Reggeon exchanges in GRT, the model includes subprocesses
with colour exchange connected to one or more Pomeron exchanges in
elastic amplitudes in the GRT. The hard gluon-gluon scattering with
large momentum transfer and the so-called semihard processes with 
quark and gluon scattering are included in the QGSM as well 
\cite{hard}. The inelastic {\it hh\/} cross section $\sigma_{in}(s)$ 
can be calculated via the real part of the eikonal $u(s,b)$ 
\beq \ds
\sigma_{in}(s) = 2\pi \int \limits_{0}^{\infty} 
\left\{ 1 - \exp\left[ - 2 u^{R}(s,b) \right]  \right\} b db \ .
\label{eq3}
\eeq
Here $s$ is the center-of-mass energy of the reaction.
The eikonal can be presented as a sum of three terms 
corresponding to soft and hard Pomeron exchange, and triple Pomeron
exchange, which is responsible for the single diffraction process,
\beq \ds
u^R(s,b) = u^R_{soft}(s,b) + u^R_{hard}(s,b) + u^R_{triple}(s,b)\ .
\label{eq4}
\eeq
Using the Abramovskii-Gribov-Kancheli (AGK) cutting rules \cite{agk} 
the inelastic cross section of {\it hh\/} interaction can be presented 
as 
\beqar \ds
\sigma_{in}(s) &=& \sum \limits_{i,j,k = 0; i+j+k \geq 1}^{ }
\sigma_{ijk}(s)\ ,\\
\sigma_{ijk}(s) &=& 2 \pi \int \limits_{0}^{\infty} b db
\exp{\left[ -2 u^R(s,b) \right]}\\
\nonumber
 &\times & \frac{\left[ 2u^R_{soft}(s,b)  \right] ^i}{i !}
\frac{\left[ 2u^R_{hard}(s,b)  \right] ^j}{j !}
\frac{\left[ 2u^R_{triple}(s,b)  \right] ^k}{k !} \ .
\label{eq5-6}
\eeqar
The last equation enables one to determine the number of cut soft and 
hard Pomerons, i.e., the number of strings and hard jets. 
The single Pomeron exchange, that can be represented by a 
cylinder-type diagram \cite{Kaid}, leads to the formation of two 
quark-diquark or quark-antiquark strings. With rising energy 
the processes with multi-Pomeron exchanges become
more and more important. The contribution of the cylinder diagrams
to the scattering amplitude increases as $s^{\alpha_P(0) -1}$, while 
that of the so-called chain diagrams corresponding to $n$-Pomeron 
exchanges $(n \geq 2)$ rises as $s^{n[\alpha_P(0) -1]}$ with 
$\alpha_P(0) > 1$ being the intercept of a Pomeranchuk pole. 
Diagrams corresponding to hard gluon-gluon scattering and double 
Pomeron exchange are shown as an example in Fig.~\ref{fig1}(a) and 
(b), respectively. An overview and recent development of the GRT can 
be found in Ref.~\cite{Wer00}. The cascade (rescattering of 
secondaries) is introduced in the QGSM as well. 
There are several other models which employ the 
colour exchange mechanism for string excitations, e.g., the dual parton
model (DPM) \cite{dpm} and the very energetic nuclear scattering
model (VENUS) \cite{venus}.

\begin{figure}[htp]
\centerline{\epsfysize=16cm \epsfbox{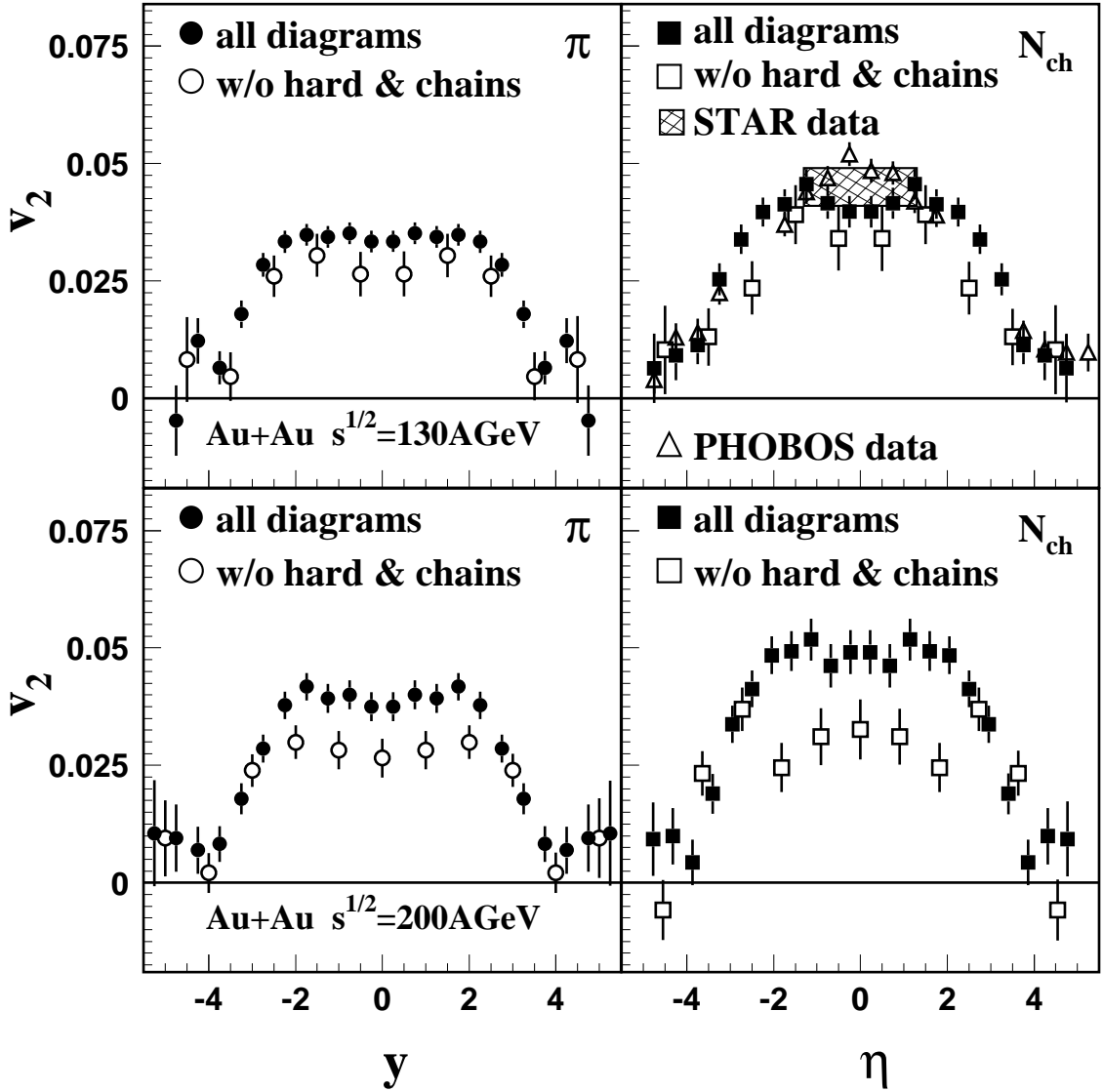}}
\vspace{0.5cm}
\caption{
Elliptic flow of pions (solid circles) and charged particles
(solid squares) as a function of rapidity $y$ and pseudorapidity
$\eta$ in minimum bias Au+Au collisions in the QGSM at 
$\sqrt{s}=130${\it A\/} GeV (upper panels) and 200{\it A\/} GeV
(lower panels). Open circles and squares show the elliptic flow in 
the model version without the hard processes and multi-chain 
diagrams. Hatched area indicates the data measured by the STAR 
Collaboration \protect\cite{ellST00}, open triangles denote the
preliminary results of the PHOBOS Collaboration 
\protect\cite{Rol01}.
}
\label{fig2}
\end{figure}

The centrality, transverse momentum, and rapidity dependence of 
anisotropic flow in the QGSM at energies from AGS to SPS has been 
studied in Refs.~\cite{Am91,Br95,flow}. Here we will focus on 
study of elliptic flow in minimum bias Au+Au collisions at two 
energies, $\sqrt{s}=130${\it A} GeV and 200{\it A} GeV, available at 
RHIC. Rapidity and pseudorapidity distributions of elliptic flow of 
pions and charges particles are depicted in Fig.~\ref{fig2}. For both 
energies elliptic flow displays strong in-plane alignment in 
accordance with the predictions of Ref.~\cite{Olli92}. In the 
mid(pseudo)rapidity the flow is almost constant. At 
$\sqrt{s}=130${\it A} GeV it rises up slightly at $|y|, |\eta| 
\approx 1.5$, and then drops with the increasing rapidity.
The mean value of the $v_2^{ch} (|\eta| < 1.3)$ equals 4.16 $\pm$ 
0.5 \%, that is very close to the value 4.5 $\pm$ 0.4 \% measured by
the STAR Collaboration \cite{ellST00}. Pseudorapidity dependence of
the elliptic flow of charged particles in the whole $\eta$ range is
also in a good agreement with the preliminary results reported by 
the PHOBOS Collaboration \cite{Rol01}. The QGSM predicts that at full
RHIC energy elliptic flow of charged particles will increase further
to $v_2^{ch} (|\eta| < 1.3) = 4.9 \pm 0.5$\%. To elaborate on the 
influence of hard processes and multi-Pomeron exchanges on the 
elliptic flow formation the flow caused by the subprocesses 
without the hard and multichain contributions is also plotted in 
Fig.~\ref{fig2}. It seems that in Au+Au collisions at 
$\sqrt{s}=130${\it A} GeV the magnitude of the signal (except of the 
midrapidity range) can be reproduced without the many-string 
processes. At $\sqrt{s}=200${\it A} GeV their role becomes more 
significant, because the elliptic flow caused by other subprocesses 
cannot exceed the limit of 3$-$3.5\% . Here it is important to
stress that the multichain diagrams alone, without rescattering,
cannot affect the elliptic flow at all. The flow increases solely
due to secondary interactions of produced particles in spatially
asymmetric system.

\begin{figure}[htp]
\centerline{\epsfysize=16cm \epsfbox{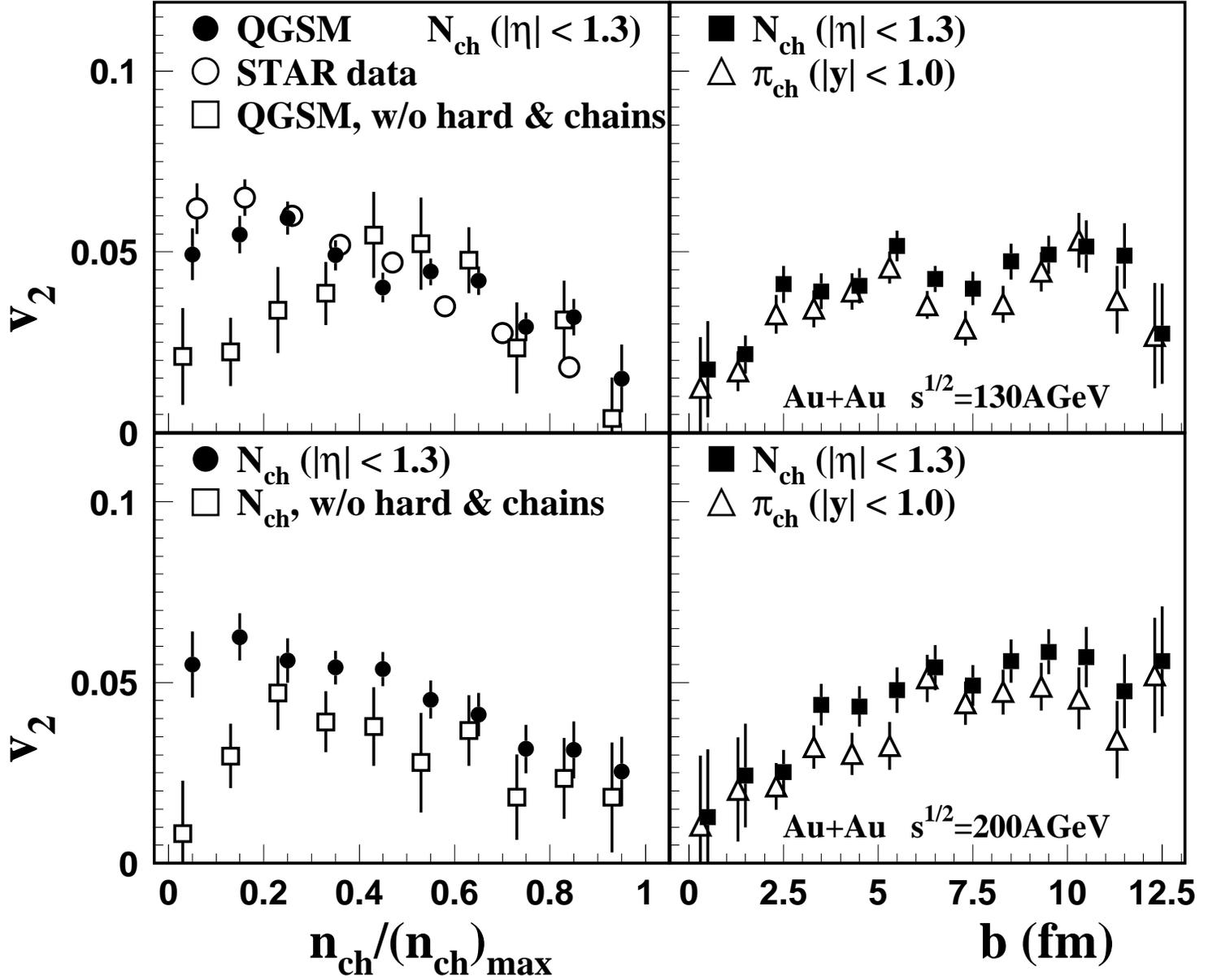}}
\vspace{0.5cm}
\caption{
Elliptic flow of charged particles (solid circles) with 
$|\eta| < 1.3$ as a function of the ratio $n_{ch}/(n_{ch})_{max}$ 
(left panels) and the impact parameter $b$ (right panels) in minimum
bias Au+Au collisions at 
$\sqrt{s}=130${\it A\/} GeV (upper row) and 200{\it A\/} GeV
(lower row). The flow of charged pions with $|y| < 1$ in the model
is shown by open triangles, open circles denote the experimental
data from Ref.\protect\cite{ellST00}. Open squares indicate the 
elliptic flow in the model version without the hard processes and 
multi-chain diagrams. 
}
\label{fig3}
\end{figure}

Figure \ref{fig3} presents the centrality dependence of elliptic
flow of charged particles. Since the centrality of events in the 
experiment has been determined via the ratio of charged particle
multiplicity to its maximum value $N_{ch} / (N_{ch})^{max}$, we 
compare the $v_2 \left[ N_{ch} / (N_{ch})^{max} \right]$ signal with 
the original impact parameter dependence $v_2(b)$.
One can see that the ratio $N_{ch} / (N_{ch})^{max}$ is a good 
criterion of the event centrality except of the very central 
region with $b \leq 2.5$ fm, where the multiplicity depends weakly
on the impact parameter. However, as a function of the impact
parameter $b$ elliptic flow is saturated at $b \approx 8$ fm for
both energies, while as a function of the multiplicity ratio 
it increases nearly linearly with decreasing multiplicity up to 
$N_{ch} / (N_{ch})^{max} \approx 0.2$. As expected, the flow in 
the midrapidity region is caused mainly by pions. The magnitude
of the pionic flow in the QGSM calculations is twice as large as 
obtained, e.g. with RQMD \cite{SPV99}. Without the many-string 
processes the QGSM is able to describe the flow only in central and 
semicentral collisions. It predicts a drop of the elliptic flow as 
the reaction becomes more peripheral, which is similar to
the predictions of other string models \cite{BlSt00,SPV99}.

The difference in centrality dependences of elliptic flow in
calculations with and without multi-string processes can be
explained as follows \footnote{This point was clarified in 
discussions with J.-Y. Ollitrault}. Hard jets and many-Pomeron 
exchanges produce more particles, thus giving rise to more 
rescatterings that drive the system toward thermal and chemical 
equilibrium. The centrality dependence
of elliptic flow in this case is very close to that obtained at
the hydrodynamic limit first in Ref.~\cite{Olli92}. In contrast,
the less number of produced secondaries would bring the system
closer to the low density limit \cite{HeLe99}, which means
incomplete (if any) thermalization of the system. In this regime 
elliptic flow in peripheral collisions vanishes much earlier 
compared to the hydrodynamic one \cite{HeLe99,VoPo00}.

\begin{figure}[htp]
\centerline{\epsfysize=16cm \epsfbox{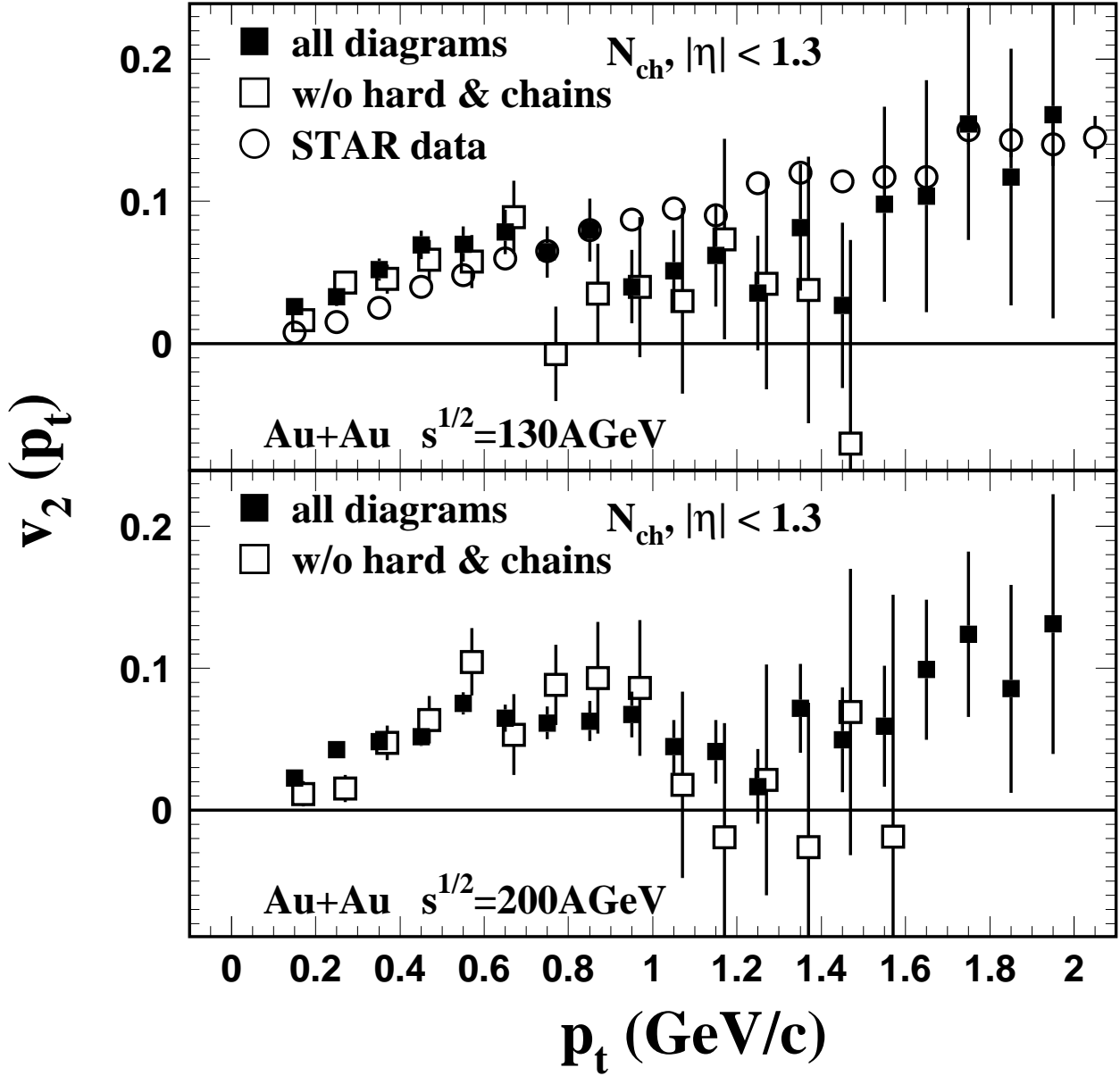}}
\vspace{0.5cm}
\caption{
Transverse momentum dependence of the elliptic flow of charged 
particles with $|\eta| < 1.3$ in minimum bias Au+Au collisions at 
130{\it A\/} GeV and 200{\it A\/} GeV. Data from 
Ref.\protect\cite{ellST00} are shown by open squares.
}
\label{fig4}
\end{figure}

The transverse momentum dependence of elliptic flow is shown in
Fig.~\ref{fig4}. The agreement with the experimental data is good, 
although the model slightly overpredicts the elliptic flow
of charged particles at $p_t \leq 0.5$ GeV/$c$. Also, the dip in
the excitation function of elliptic flow at $p_t \approx 1$~GeV/$c$ 
looks peculiar. Nevertheless, both effects are
attributed to the lack of heavy resonances in the model. This, 
however, cannot affect the rapidity and centrality distributions
where the spectra of particles are integrated over the whole 
transverse momentum range. Similar to the centrality dependence, the
calculations without the many-string processes can explain the 
experimental $v_2(p_t)$ data only at $p_t \leq 0.8$~GeV/$c$. To 
describe the further increase of elliptic flow with rising $p_t$ the 
full set of diagrams is required. At $p_t \geq 1.8$~GeV/$c$ elliptic 
flow saturates in accord with experimental results \cite{Sne01}.
Note that even without the many-string processes the elliptic flow of 
hadrons in the QGSM is stronger than that of the string models based 
on the FRITIOF routine. This is due to the fact that 
the string stretched between the constituents belonging to different
hadrons is not parallel to the beam axis, thus giving the additional
transverse push to secondaries.

In conclusion, the microscopic quark-gluon string model based on the 
colour exchange mechanism is able to reproduce quantitatively 
experimental data on elliptic flow in Au+Au collisions at RHIC 
energies  quite well, i.e. the microscopic models are still on the 
market for ultrarelativistic heavy-ion collisions. 
The key processes in the formation of strong elliptic flow at collider 
energies appear to be hard Pomeron exchange, resulting in the creation 
of hard partonic jets, and soft multi-Pomeron exchanges, accompanied
by the subsequent rescattering of secondaries in spatially asymmetric
system. Note, that directed and elliptic flow strongly depends on the
mean free path of hadrons, which is inversely proportional to the 
particle density and interaction cross section. The larger the 
density (or cross section) of particles participating in secondary 
rescattering, the larger the elliptic flow \cite{MoGy01}.
Therefore, multi-Pomeron exchanges and hard processes increase the
yield of secondaries at the initial stage of the reaction thus 
enhancing the elliptic flow.
Without the many-string processes the model cannot describe the rise
of elliptic flow $v_2(b)$ for peripheral events, while the magnitude of 
the flow $v_2^{ch}(y)$ in Au+Au collisions at 130{\it A\/} GeV and 
200{\it A\/} GeV cannot exceed the limit of 3.5\%.
For the $v_2(p_t)$ distribution the QGSM predicts the saturation of
elliptic flow at $p_t \geq 1.8$~GeV/$c$. This finding agrees
quantitatively with the experimental results and qualitatively with
the predictions of other string models, but deviates from the
hydrodynamic calculations. More comparison is needed to distinguish
unambiguously between hydro- and string- approaches, and to find
clear signals of the quark-gluon plasma formation.

{\bf Acknowledgements}. 
We are indebted to L.~Csernai, J.-Y.~Ollitrault,
D.~Strot\-tman, V.D.~Toneev, and S.A.~Voloshin for interesting 
discussions and fruitful comments.
This work was supported in part by the Bundesministerium f\"ur 
Bildung und Forschung (BMBF) under contract 06T\"U986, and by the
Bergen Computational Physics Laboratory (BCPL) in the framework of the
European Community - Access to Research Infrastructure action of the
Improving Human Potential Programme.

\newpage

\end{document}